# Performance Evaluation: Ball-Tree and KD-Tree in the context of MST

Hazarath Munaga[1,1], Venkata Jarugumalli[1]

[1] Dept. of Information Technology
University College of Engineering
JNTUK, A.P. India
{hazarath.munaga, venkata.jarugumalli}@gmail.com

**Abstract.** Now a day's many algorithms are invented / being inventing to find the solution for Euclidean Minimum Spanning Tree (*EMST*) problem, as its applicability is increasing in much wide range of fields containing spatial / spatio – temporal data viz. astronomy which consists of millions of spatial data. To solve this problem, we are presenting a technique by adopting the dual tree algorithm for finding efficient EMST and experimented on a variety of real time and synthetic datasets. This paper presents the observed experimental observations and the efficiency of the dual tree framework, in the context of kd-tree and ball-tree on spatial datasets of different dimensions.

## 1 Introduction

Minimum Spanning Tree (*hereinafter*, MST) is the one of the oldest and most thoroughly studied problem in computational geometry [1]. The Minimum-weight Spanning Tree or simply MST problem is one of the well-known optimization problems for finding minimum weighted spanning tree for both undirected and directed graphs. MST have many applications in computer design, communication, transportation and some other problems that can apply directly/indirectly are wireless network connectivity, clustering and classification of different data namely spatial data. In literature, several authors proposed a variety of greedy algorithms [2, 3, 4]. The general drawback with these greedy algorithms is that these cannot handle large amount of data and performance bottleneck will happened and later some more advanced algorithms are developed to solve this problem [5]. In this paper, authors evaluated the performance of dual tree algorithmic framework [6] using single linkage clustering [7] and compared the performance of the dual tree framework in the context of kd-tree and ball-tree. As per the Ref. [8] theoretically single-linkage clustering provides optimal clusters.

---

[1] *Dr MHM Krishna Prasad, Associate Professor & Head*

## 2 Related Work

Many of the MST Algorithms are utilized Tarjan's Blue rule [10] to find the minimum weight edge. Using this rule many of the greedy algorithms are developed viz. Prims with $O(m+nlogn)$ and Krushkal with $O(mlogn)$ time complexity's respectively. In Dual-Tree Boruvka's Algorithm, the efficiency of the Boruvka's algorithm mainly depends on the adopted nearest neighbor technique. If we find nearest neighbors efficiently by using some intelligent data structure like spatial trees, kd-tree, ball tree or any spatial tree, the performance of the Boruvka's algorithm can be enhanced. For example, for the query set $Q$ and reference set $R$, then for each point $q \in Q$ and $r \in R$ such that $d(q, r)$ must be minimized and call it as nearest neighbor pair. Native approach will take $O(n^2)$ run time for finding nearest neighbor pair for $n$ number of components[11]. Hence, in this paper, authors adopted the algorithms proposed by [7] to evaluate the performance of Dual Tree Boruvka and compared with kd-tree and ball-tree.

**KD-tree** [12] is one of the space partitioning tree for organizing k-dimensional data points. For building this kd-tree of n points it takes $O(n\ log\ n)$ if we use the linear median finding algorithm described by [13] and for adding new point to the balanced tree it takes $O(log\ n)$ and for removing a point from tree takes $O(log\ n)$, because of removing a point from tree the balance of the tree will change then we need to rebalance the tree so for both adding and removing point from tree will take same time complexity of $O(log\ n)$.

**Ball Tree** [14] is also binary tree data structure for maintaining spatial data hierarchically like kd-tree and oct-tree [15]. Each node in the ball-tree referred as ball contains a region of Euclidean points bounded by a hyper-sphere and interior balls are small containing their children balls. One can specify the ball $n+1$ float values as co-ordinates and $r$ radius as its center. Like kd-tree the ball tree also uses top down approach for building the tree recursively from top to down by choosing the split dimension and splitting value to find these values, balls are sorted along each dimension and store the cost in an array. Best dimension and split location can be found in $O(nlog(n))$, so, time complexity to construct the ball-tree is $O(n(log\ n)^2)$.

## 3 Implementation

In this paper, authors adopted Dual-tree Boruvka in the context of kd-tree and ball-tree for finding the EMST. Experiments are performed on synthetic datasets and real time datasets [9]. Both algorithms are implemented and incorporated in weka3.6.4 [16] and compiled with the jdk1.6. Following Table and Fig. 1 shows the observations.

**Table-1** Performance of *kd-tree* and *ball-tree* for 50 dimensions

| Operation Instances | Build | | Insertion | | Deletion | | N-NN Search | |
|---|---|---|---|---|---|---|---|---|
| | kd-tree | ball-tree | kd-tree | ball-tree | kd-tree | ball-tree | kd-tree | ball-tree |
| 10000 | 187 | 1203 | 188 | 1187 | 218 | 1172 | 31 | 47 |
| 25000 | 547 | 3469 | 531 | 3468 | 516 | 3454 | 96 | 110 |
| 50000 | 1344 | 7906 | 1172 | 8094 | 1156 | 8093 | 118 | 219 |
| 100000 | 2578 | 16547 | 2750 | 16437 | 2812 | 16485 | 391 | 422 |
| 200000 | 5937 | 36672 | 5688 | 36703 | 5640 | 36375 | 875 | 907 |

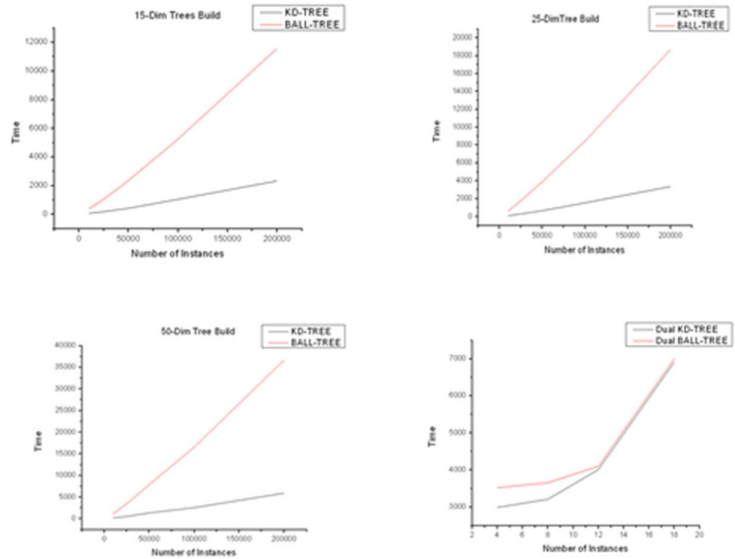

**Fig-1**: (a) time taken for building kd-tree and ball-tree for 15-dimensions (b) time taken for building kd-tree and ball-tree for For 25-dimensions (c) time taken for building kd-tree and ball-tree for 50-dimensions of data; (d) Comparison of dual tree algorithm using both kd-tree and ball-tree

From the above results, one can easily observe that the computational time for constructing ball-tree takes more time than the kd-tree, and as the dimensionality increases the performance of kd-tree is also increases when compared with ball-tree. Figure-1(d) shows the performance of DualTree Boruvka using kd-tree as well as ball-tree for varying dimensions on real-time datasets obtained from SDSS. Both algorithmic performance is nearly equal but overall performance differs with creation of tree structure, i.e., kd-tree takes $O(n \log n)$ and ball-tree takes $O(N(\log N)^2)$ for n points in Euclidean space. Finally, from the experimental results, authors conclude that DualTree Boruvka on kd-tree performs faster than the DualTree Boruvka on ball-tree for finding Euclidean MST.

## 4   Conclusion

In this paper, authors compared kd-tree and ball-tree based dual tree Boruvka algorithm for finding Euclidean Minimum Spanning Tree (*EMST*). For finding efficient EMST, authors adopted dual tree algorithm and experimented on a variety of real time and synthetic datasets of various dimensions. From the experimental observation, authors conclude that the *kd-tree performs faster than the ball-tree for not only constructing the tree and also for solving the EMST problem. Moreover, the kd-tree based dual tree Boruvka is giving good results than the ball-tree based dual-tree Boruvka.*

## References

1. Preparata, F., Shamos, M.: Computational Geometry. Springer-Verlag, New York (1985).
2. Nesetril.: Otakar Boruvka on minimum spanning tree problem Translation of both the 1926 papers, comments, history. Discrete Math., vol.233, pp.3–36, (2001).
3. Prim, R. C.: Shortest connection networks and some generalizations. J. Bell Sys. Tech., 1389—1401(1957).
4. Kruskal, J. B.: On the shortest spanning subtree of a graph and the traveling salesman problem. Proc. Am. Math. Soc., 7:48 -- 50 (1956).
5. Narasimhan, G., M. Zachariasen, M., Zhu, J.: Experiments with computing geometric minimum spanning trees. In: Proceedings of ALENEX'00, pp.183--196 (2000).
6. Gray, A., Moore, A., W.: N-body problems in statistical learning. In: Advances in Neural Information Processing Systems, pp.521—527, (2001).
7. William, B. M., Parikshit, R., Alexander G.: Fast Euclidean Minimum Spanning Tree: algorithm, analysis, and applications. In: 16th ACM SIGKDD International conference on Knowledge discovery and data mining, pp. 603-612, (2010).
8. Balcan, M., Blum, A., Vempala, S.: A discriminative framework for clustering via similarity functions. In: 08 Proceedings of the 40th annual ACM symposium on Theory of computing, pp. 671--680. ACM, New York (2008).
9. Sloan Digital Sky Survey, http://sdss2.lib.uchicago.edu/dr7/en/, accessed on 15[th] Jan 2011
10. Tarjan, R. E: Data Structures and Network Algorithms. In: Society for industrial Applied Mathematics, vol. 44 (1983).
11. Graham, R.L., Pavol, H,: On the history of the Minimum Spanning Tree Problem. Annals of History of Computing. J. IEEE Ann. Hist. Comput. 7, 43—57, (1985).
12. Moore, A.,W.: An intoductory tutorial on kd-trees. Technical Report No. 209, Computer Laboratory, University of Cambridge (1991).
13. Cormen, T. H., Leiserson, C. E., Rivest, R. L, Clifford S.: Introduction to Algorithms 3[rd] edition. MIT Press and McGraw-Hill, (2009).
14. Omohundro, S., M.: Five Balltree Construction Algorithms, ICSI Technical Report TR-89-063 (December 1989)
15. Warren, M., S., Salmon, J., K.: A parallel hashed Oct-Tree N-body algorithm. In: Proceedings of the ACM/IEEE conference on Supercomputing, pp. 12--21, (1993).
16. Machine Learning Group at university of Waikato, http://www.cs.waikato.ac.nz/ml/weka/, accessed on 25[th] August 2010.